# Forecasting stock market returns over multiple time horizons

Dimitri Kroujiline*[1], Maxim Gusev[2], Dmitry Ushanov[3], Sergey V. Sharov[4] and Boris Govorkov[2]

## Abstract

In this paper we seek to demonstrate the predictability of stock market returns and explain the nature of this return predictability. To this end, we introduce investors with different investment horizons into the news-driven, analytic, agent-based market model developed in Gusev et al. (2015). This heterogeneous framework enables us to capture dynamics at multiple timescales, expanding the model's applications and improving precision. We study the heterogeneous model theoretically and empirically to highlight essential mechanisms underlying certain market behaviors, such as transitions between bull- and bear markets and the self-similar behavior of price changes. Most importantly, we apply this model to show that the stock market is nearly efficient on intraday timescales, adjusting quickly to incoming news, but becomes inefficient on longer timescales, where news may have a long-lasting nonlinear impact on dynamics, attributable to a feedback mechanism acting over these horizons. Then, using the model, we design algorithmic strategies that utilize news flow, quantified and measured, as the only input to trade on market return forecasts over multiple horizons, from days to months. The backtested results suggest that the return is predictable to the extent that successful trading strategies can be constructed to harness this predictability.

[1] LGT Capital Partners, Pfäffikon, Switzerland, dimitri.kroujiline@lgt.com (*contact author). [2] IBC Quantitative Strategies, Tärnaby, Sweden. [3] Moscow State University, Department of Mechanics and Mathematics, Moscow, Russia. [4] Lobachevsky State University, Advanced School of General & Applied Physics, N. Novgorod, Russia.

**Introduction**

The integration of news analytics into trading strategies continues to be at the R&D forefront in the investment industry. The efforts have mainly focused on constructing early indicators for a change in investor sentiment to enable a trader to act ahead of the majority of investors. The potential for success is therefore reliant on the speed and precision with which information retrieval and text parsing algorithms process vast amounts of data and recognize, in the incoming flow of news, the events that may move prices substantially.

As prices generally tend to quickly reflect new information, the objective of news-based trading is to produce short-term (intraday) strategies. On short timescales prices may be assumed to react to news linearly. This linearization is helpful because a trade's sign would then depend only on the sign of sentiment assigned to a news item and its size on price sensitivity – each still being a formidable task. As with any other short-term trading strategy, the downside here is a limited capacity and high, turnover-driven costs. Furthermore, should successful strategies eventually be developed – so far results have been mixed to the best of our knowledge – competition in this segment will spur an "arms race" of speed, leading to increasingly fast price reaction to news, exacerbating capacity restrictions and reducing profit margins.[1]

The above-described approach is based on a premise that financial markets need a finite amount of time to digest news. In other words, return prediction using news analytics relies on the market's informational inefficiency in the time interval where prices adjust to new information. Does it then follow that news analytics are necessarily useless for trading over longer horizons?

---

[1] For example, competition among high frequency trading firms has increased the trade execution speed from roughly 100 milliseconds to 10 microseconds over the last decade.

At first glance, the answer seems to be a firm "yes". Indeed, while it appears reasonable to suppose that a market can "remember", in terms of its reaction, previous events on the order of minutes and hours, this same supposition sounds absurd when considering periods that span days, weeks or months. Yet, there is certain empirical evidence of long-term return predictability.[2]

It might be useful to tackle the problem from a different angle and consider whether there exist any logical possibilities for long-term predictability. Incidentally, such a possibility does exist, provided the mechanism of price formation over long time horizons is different from that involved on short timescales. Let us hypothesize how it may operate.

Efficient markets ensure that new information is manifested in a change in market price shortly following its release. However, this price change can also be an important event on its own that will draw media response, which can incite subsequent price changes, causing in turn further news releases and so forth. Thus, the original event may trigger a "ripple effect" of the interlinked price changes and news releases, unfolding over an extended time period. This implies that news can have a long-lasting impact in the market, arising through this feedback mechanism,[3] which is absent on short timescales as we will see later.

---

[2] There is a large body of research that examines, typically applying regression methods, the predictive power of observable variables, such as the dividend yield and many others; see, for example, Fama and French (1988, 1989), Campbell and Shiller (1988a,b), Baker and Wurgler (2000), Campbell and Thompson (2008), Cochrane (2008). However, the evidence for return prediction remains inconclusive: e.g. Ferson et al. (2003, 2008), Goyal and Welch (2003, 2008). The model of market dynamics that we develop here is fundamentally nonlinear, indicating that causal relations among the variables are substantially more complex than regression dependence. It follows that the standard approach to return prediction, based on regression methods, may be ill-suited to capture this predictability; e.g. Novy-Marx (2014) vividly pointed out this limitation by extending stock market predictive regressions to a number of rather implausible variables, such as sunspot activity and planetary motion. We propose an alternative approach that combines theoretical models with empirical data to explore whether stock market returns can be predicted in an economically significant manner.

Thus, from a news-price system with no feedback, likely valid on the order of a day or less, we move toward a mutually-coupled news-price system, operating over longer horizons. Consequently, we must apply a different framework for return prediction. Whereas the short-term prediction requires fast detection of the news releases that may provoke material changes in price, the long-term prediction, on the contrary, can only be based on the regularity of the system's behavior. In other words, we must develop a dynamic model that correctly describes interaction between news and price. Then, provided the model admits non-stochastic solutions, it would be enough to know the market position in the news-price reference frame to forecast return by following the market evolution path provided by the model.

Gusev et al. (2015) proposed such a model, which describes stock market dynamics in terms of the interaction between prices, opinions and information. The model was formulated as an Ising-family agent-based model[4] with two types of interacting agents: investors, who invest or divest according to their opinions, and analysts, who interpret news, form opinions and channel them to investors.[5] To derive the model equations in analytic form and facilitate its study, it was assumed that investors made up a homogeneous group in which any two market participants interacted identically. Despite this and other simplifications, the model reproduced the price path and return distribution of the S&P 500 Index within reasonable tolerance. Based on these results, the authors suggested that stock market returns are predictable, but did not conduct tests of this predictability as their model was still too coarse-grained to produce sufficiently precise forecasts.

---

[3] The idea that the observations of price changes may generate a feedback loop that significantly affects market dynamics is not new (see a review by Shiller, 2003). However, its application for return forecasting, which is the subject of the present work, is nontrivial due to nonlinear behaviors induced by it.

[4] A family of models, named after Ising (1925), developed originally to explain the phenomenon of ferromagnetism via the interaction of discrete atomic spins in an external magnetic field and later broadly applied to study problems in social and economic dynamics (see reviews by Castellano et al., 2009; Lux, 2009;

The main objective of the present work is to design a model with improved forecast precision and apply it to demonstrate that returns can be predicted over time horizons longer than one day. To accomplish this, we introduce heterogeneity into Gusev's et al. (2015) framework by replacing homogeneous investors with groups of investors that have different investment horizons. This enables us to extract characteristic dynamics on different timescales and produce market forecasts of multiple time horizons, upon which we construct trading strategies.

---

Slanina, 2014; Sornette, 2014). We take note of two recent works that share common ground with Gusev et al. (2015). First, Franke (2014), referencing Lux's (1995) analytic stock market model, studied a generic sentiment-driven economic model with feedback, which has some features similar to those found in Gusev et al. (2015). Second, Carro et al. (2015) investigated the influence of exogenous information on sentiment dynamics in the stock market, which is also a central theme in Gusev et al. (2015).

[5] This approach contrasts with that of the established agent-based financial models, where market dynamics are sought to emerge, primarily, through the interaction among agents pursuing different trading strategies, such as the influential work by Lux and Marchesi (1999) among many others.

Owing to this heterogeneity, the model achieves a greater explanatory capacity that serves to highlight certain market behaviors. For example, the familiar self-similar patterns in price returns on different timescales can be explained as a result of sentiment dynamics synchronization across different investor groups (Sections 2.1-2.3); the transition between bull- and bear markets is shown to occur as a cascade, whereby groups with longer investment horizons follow groups with shorter horizons from one sentiment equilibrium to the other (Sections 2.1-2.3); and it is demonstrated that the negligible impact of price feedback in the intraday dynamics allows sentiment to quickly adjust to incoming news, thereby imposing informational efficiency on these timescales (Section 2.4).

The paper is organized as follows. Section 1 describes the news-driven market model with homogeneous investors and develops the model with heterogeneous investors. Section 2 studies the heterogeneous model analytically, numerically and empirically. Additionally, Section 2.4 examines market efficiency on intraday timescales. Section 3 lays out the design and backtests the trading strategies. Section 4 further discusses the nature of return predictability. Section 5 provides a summary of conclusions.

## 1. Models

This section introduces the model with homogeneous investors, developed in Gusev et al. (2015), and using it as a starting point, derives a more general model with heterogeneous investors that we will apply for market forecasting.

### 1.1 Homogeneous model

The model of stock market dynamics in Gusev et al. (2015) is formulated as a dynamical system governing the evolution of three independent variables: market price $p$, investor sentiment $s$ and information flow $h$. It was obtained by defining, based on observed behaviors, interactions among agents at a micro level and applying methods from statistical mechanics to produce dynamic equations for averaged variables at a macro level. Before exploring the equations, it may be helpful to explain the proper context in which sentiment and information are used in the model.

Investor sentiment is defined as a summary view on future market performance, averaged across the investment community, and is determined as the ratio of the number of investors who opine that the market will rise minus the number of investors who opine that the market will fall over the total number of investors. Thus, sentiment $s$ can vary between -1 and 1. By this definition sentiment $s$ encompasses all types of opinions, irrespective of whether an opinion has been formed rationally or irrationally.

Information flow considered in the model as relevant comprises publicly expressed opinions about the direction of anticipated market movement. It is quantified similarly to sentiment as the ratio of the number of news items with positive expectations minus the number of news items with negative expectations over the total number of relevant news items. Like sentiment, information $h$ is bounded between -1 and 1. The fact that $h$ can be readily measured allows the model to be empirically verified.[6]

---

[6] Extensive research has been done on empirical measures of sentiment – which include indices based on periodic surveys of investor opinion; various proxies such as trading volume, call vs. put contracts and others; and applications of machine learning and rule-based techniques for parsing financial news and social media content – and on their correlation with price movement (e.g. Antweiler and Frank, 2004; Brown and Cliff, 2004; Baker and Wurgler, 2007; Das and Chen, 2007; Tetlock, 2007; Loughran and McDonald, 2011; Lux, 2011; Da et al., 2014). Alternatively, Gusev et al. (2015) suggested a rule-based parsing methodology for measuring $h$ and calculated sentiment $s(h)$ from this empirical $h$, using the homogeneous model described in this section.

The model is described by the differential equations:[7]

$$\dot{p} = a_1 \dot{s} + a_2(s - s_*), \tag{1a}$$

$$\tau_s \dot{s} = -s + \tanh(\beta_1 s + \beta_2 h), \tag{1b}$$

$$\tau_h \dot{h} = -h + \tanh(\kappa_1 \dot{p} + \kappa_2 \xi_t). \tag{1c}$$

The first equation, derived by observing that investors tend to act on their opinions differently over short and long horizons, establishes a phenomenological relation between the change in log price and investor sentiment. This equation states that price changes proportionally, first, to the change in sentiment and, second, to the deviation of sentiment from a certain reference level $s_*$. The former is the main source of short-term price variation, while the latter determines leading behavior over long-term horizons.

The second and third equations were derived together as a single system, using methods from statistical mechanics. The second equation describes the change in sentiment due to the impact of information flow on investors via the term $\beta_2 h$ and the interaction among investors via the term $\beta_1 s$, where $\tau_s$ is the characteristic time of sentiment variation and $\beta_1$ determines the relative importance of the herding behavior and the random behavior of investors. Information flow acts as a force that moves sentiment away from equilibrium. If it were to cease, sentiment would come to rest at a nonzero value for $\beta_1 > 1$ (ordered state) or at zero for $\beta_1 < 1$ (disordered state).

---

[7] Gusev et al. (2015) (Eq. 13, Fig. 12). The dot denotes the derivative with respect to time. Parameters $a_1, a_2, \beta_1, \beta_2, \kappa_1, \kappa_2, \tau_s, \tau_h$ are positive, while $s_*$ can take any sign. The parameter values, estimated using the empirical data, are provided in Table 1 of that same paper. We note that equation (1b), with $h$ as an exogenous variable, was originally obtained by Suzuki and Kubo (1968) in the context of a purely statistical mechanics problem.

The third equation states that the change in information flow is caused by exogenous news $\xi_t$ and news about price changes $\dot{p}$, with $\tau_h$ being the characteristic response time.[8]

Equations (1) form a three-dimensional nonlinear dynamical system. Each point in the phase space $(h, s, p)$ represents a unique market state and each solution $(h(t), s(t), p(t))$ represents a phase trajectory of market evolution. This evolution is driven by the flow of exogenous news $\xi_t$ that induces random fluctuations of the phase trajectory and by the feedback mechanism $h \to s \to p \to h$ that generates inertial dynamics, giving rise to deterministic behaviors. It follows that according to this model, market evolution may contain deterministic regimes and thus be potentially predictable.

Equations (1) were obtained under a simplifying assumption of the all-to-all interaction pattern among agents.[9] In reality, interaction among investors is hardly so simple. The utilization of more sophisticated patterns of interaction in the Ising-type models is known to cause the emergence of heterogeneous structures – in the present case, clusters of investors with co-aligned sentiments. Because the size of a cluster determines its reaction time to incoming information, this heterogeneity can generate diverse dynamics involving interactions on many timescales. It follows that model (1) should be regarded as a coarse-grained approximation that determines the average investor behavior evolving within a single timeframe $\tau_s$[10]. Thus, although this model provides certain valuable insights into market dynamics, it is insufficiently realistic for market forecasting (especially over periods shorter than $\tau_s$). We must refine this model to improve precision, which is the subject of the next section.

---

[8] This form of the equation neglects the impact of direct interaction between the agents, omitting the terms proportional to $h$ and $s$ in the argument of the hyperbolic tangent (Gusev et al., 2015: Eq. 12c).

[9] The all-to-all interaction mode is the leading-order approximation for a general interaction topology in this model.

**1.2 Heterogeneous model**

We wish to replace the above-described framework, where each investor interacts with all other investors with the same strength, by a framework with a more realistic interaction pattern. Selecting such a pattern in the form of rules applicable to individual agents is a hard problem to solve. This is because there exist many plausible choices for interactions at the micro-level and the model's statistical properties will be sensitive to these choices.[11] Additionally, it would be difficult, if at all possible, to derive a closed-form dynamical system for the evolution of macro-level variables based on the interaction patterns more complex than all-to-all.

Instead, it may be more practical to account for investor heterogeneity via a phenomenological approach. As discussed above, we expect that realistic interaction patterns would produce clusters of investors, each characterized by a specific (as a function of size) response time to incoming information, which we can assume proportional to the investment time horizon. It is therefore sensible to select the investment horizon as the attribute whereby variability is introduced into the model. Hence, we wish to modify model (1) by populating it with the investors that have various investment horizons.[12]

---

[10] Gusev et al. (2015) estimated $\tau_s$ to be around one month.

[11] Cont and Bouchaud (2000) addressed heterogeneity in opinion formation as a topological problem within the framework of percolation theory from chemistry and physics, leading to the emergence of clusters of investors with shared sentiment. Following this work, a number of percolation models have been proposed that replicate some of distinctive market behaviors. However, as mentioned above, the results are sensitive to the choice of topology in a model and it is difficult to economically justify any one particular topology choice.

Let us make assumptions on how these investors would interact. Presumably, any organization, whether in the investment industry or elsewhere, should tend to connect best with its peers, owing to shared professional interests. For example, long-term investors, such as pension plans, have little in common with wealth management companies oriented toward mid-term performance and even less so with the day-trading community. Each of these investment industry segments maintains its own professional publications, conferences, seminars, awards and other platforms for discourse that promote networking and interaction. Therefore, we can suppose that interaction within the networks of peers, whom we propose to identify with respect to the investment horizon, is more efficient than across them.

We have thus arrived at a framework where investors with similar horizons form peer networks or groups within which they interact efficiently, but have little interaction externally, at the same time being impacted in equal measure by information flow $h$. In the limiting case, we can assume, first, that the interaction pattern within each peer group is all-to-all and, second, that there is no interaction across these groups. This enables us to apply equations (1b,c) to describe the market with $N$ participating peer groups as follows:

---

[12] Similarly, the average memory timespan can be presumed proportional to the investment time horizon. We note that the agent-based market model developed in a series of publications by Levy, Levy and Solomon (see Levy et al., 2000) includes investors with different memory timespans.

$$\tau_i \dot{s}_i = -s_i + \tanh(\beta_1 s_i + \beta_2 h), \quad i = 1, 2, \dots, N, \tag{2a}$$

$$\tau_h \dot{h} = -h + \tanh(\kappa_1 \dot{p} + \kappa_2 \xi_t), \tag{2b}$$

where $s_i$ is the average sentiment of the $i$-th group, which having been normalized by the total number of investors in the group takes values between -1 and 1, and $\tau_i$ is its investment horizon. Note that both the "herding" parameter $\beta_1$ and the constant $\beta_2$, which determines sensitivity to information flow, are assumed to be in the leading order uniform across the groups.

Next, we must find the relation between $s_i$ and market price. The sentiments $s_i$ sum to yield the overall market sentiment $s$, given in a general form by $s = \sum \alpha_i s_i$. According to equation (1a), the market sentiment influences the evolution of the market price, so that the (unknown) weights $\alpha_i$ determine the relative contributions of different investor groups to price movements. On the other hand, the price actually changes in response to capital flows. Therefore, we can try to deduce $\alpha_i$ by considering how capital flows are triggered by different investor groups acting on their sentiments.

We note that Gusev et al. (2015) derived equation (1a) by proposing that capital flows in the market have different causes on different timescales, i.e. being driven by $ds$ at $t \ll \tau_s$ and by $s - s_*$ at $t \gg \tau_s$, where $\tau_s$ has the meaning of the average memory timespan or the average investment horizon.[13] Let us now consider the $i$-th network of investors, characterized by the horizon $\tau_i$, on the timescales $t$ where $t \ll \tau_i$. In this regime, $ds_i$ would cause the capital flow $dc_i \sim c_i ds_i$, where $c_i$ is the total capital managed within this network.

---

[13] The basic argument is as follows: Consider an investor who has just allocated capital to the market. The following day, this investor is unlikely to amend her allocation unless her sentiment has changed. This is because the capital that the investor has already deployed reflects this same level of sentiment. Therefore, ignoring external constraints, investment allocations can be presumed to be driven by the change in sentiment on timescales where the investors' memory of past sentiment levels persists ($t \ll \tau_s$). Conversely, on longer

To estimate $c_i$, we note that all else being equal, a short-term (high-turnover) investment strategy has a lower capital capacity due to liquidity constraints than a long-term (low-turnover) strategy. In other words, liquidity constraints effectively cap $c_i$ on different investment horizons. It is then reasonable to suppose that $c_i$ is generally an increasing function of $\tau_i$, i.e. the longer the horizon the more capital can on average be managed with no material price impact.

As a first approximation, we can assume $c_i$ to be proportional to $\tau_i$. This choice is supported by the following observation. The maximum capital amount that can be invested or divested per unit of time without materially affecting price is determined by instantaneous liquidity and so, in the context of our framework, must be the same for each investor group, independent of its investment horizon $\tau_i$ and capital under management $c_i$. This amount is given by $dc_i \sim c_i ds_i$ per $dt$ or by $\dot{c}_i \sim c_i \dot{s}_i$ in the limit $dt \to 0$. Therefore, we expect $\dot{c}_i \sim \dot{c}_j$ or $c_i \dot{s}_i \sim c_j \dot{s}_j$ for any $i$ and $j$. This can be satisfied only if $c_i \sim \tau_i$ because $\dot{s}_i \sim 1/\tau_i$ as follows from (2a).

Thus, we set $c_i \sim \tau_i$ which yields $dc_i \sim \tau_i ds_i$ and $\dot{c}_i \sim \tau_i \dot{s}_i$. As described above, $\dot{c}_i \sim 1$ since $\dot{s}_i \sim 1/\tau_i$. The implication is that the longer a group's horizon, the more money it manages, but the slower its sentiment varies, so that all groups cause on average comparable capital flows per unit of time. This

---

timescales ($t \gg \tau_s$) investors would invest or divest depending on the level of sentiment itself since their previous allocation decisions would not be linked in their memory to the previous levels of sentiment. Since capital flows lead to price changes, these two asymptotic views can be superposed to yield the approximate price-sentiment relation in the form of equation (1a). For details, see Gusev et al. (2015) (Section 1.3.1).

is sensible because otherwise some investors would systematically dominate others, which is not observed in liquid markets.

To conclude, we assume $c_i \sim \tau_i$ to obtain $\dot{c}_i \sim \tau_i \dot{s}_i$ at $t \ll \tau_i$. Similarly, we obtain $\dot{c}_i \sim \tau_i(s_i - s_{*i})$ at $t \gg \tau_i$. We superpose the asymptotic relations $\dot{c}_i \sim \tau_i \dot{s}_i$ and $\dot{c}_i \sim \tau_i(s_i - s_{*i})$ and because the change in market price is determined by the net flow of capital into or out of the market, sum across all $i$ to derive the approximate equation of price formation as a function of sentiment:

$$\dot{p} = a \sum \dot{c}_i = a_1 \left( \frac{\sum \tau_i \dot{s}_i}{\sum \tau_i} \right) + a_2 \left( \frac{\sum \tau_i (s_i - s_{*i})}{\sum \tau_i} \right) = a_1 \dot{s} + a_2 (s - s_*), \qquad (3)$$

where the constants $a_1$ and $a_2$ are positive and the constant $s_*$ can be of any sign. Equation (3) contains the expressions for the weights $\alpha_i$, i.e. $\alpha_i = \tau_i / \sum \tau_i$, implying that the sentiments $s_i$ of the networks with various $\tau_i$ and the overall sentiment $s$ in the market are related by the formula:[14]

$$s = \frac{\sum \tau_i s_i}{\sum \tau_i}. \qquad (4)$$

The heterogeneous market model is then given by the dynamical system:

---

[14] This equation should be treated as an average relation applicable under normal market conditions or over extended time periods. In particular, it is not expected to hold during spikes in trading activity, such as those accompanying market crashes. Additionally, it may not be true for groups with very long investment horizons because on the corresponding timescales effects due to the finite size of the market can affect the assumed linear dependence between the horizon and the amount of investment capital. Nevertheless, as a first approximation, this relation will prove helpful for gaining insight into market dynamics on the relevant timescales.

$$\dot{p} = a_1 \dot{s} + a_2(s - s_*), \qquad (5a)$$

$$\tau_i \dot{s}_i = -s_i + \tanh(\beta_1 s_i + \beta_2 h), \quad i = 1, 2, \dots, N, \qquad (5b)$$

$$\tau_h \dot{h} = -h + \tanh(\kappa_1 \dot{p} + \kappa_2 \xi_t), \qquad (5c)$$

where the aggregate sentiment *s* is defined by (4).

According to this dynamical system, investor groups with different investment horizons collectively form the aggregate sentiment that determines the market price, which in turn influences the information flow that acts on all groups participating in the market. Although there is no direct interaction among the groups, each continues to impact the others by eventually contributing to the common information flow. Thus, the information flow plays a dual role: it is a force that impacts the sentiments of different investor groups and also a link through which these sentiments are mutually coupled.

Setting $N = 1$ in equations (5) recovers equations (1), so that the heterogeneous model studied here encompasses, as a particular case, the homogeneous model of Gusev et al. (2015). Being more general, system (5) possesses richer dynamics and reveals new behaviors stemming from the above-described indirect interaction among the investor groups, which we will inspect in Section 2.

## 2. Study of heterogeneous model

This section studies the model with heterogeneous investors. Section 2.1 offers a preliminary analysis of the main effects expected in this model. Sections 2.2 and 2.3 investigate the model numerically, using direct simulations and empirical data, respectively. Section 2.4 applies the model to demonstrate that the efficient market regime occurs on short timescales. The relevant technical details are in Appendix A.

## 2.1. Preliminary analysis: key effects

We can substitute $\dot{p}$ from (5a) into (5c) to obtain a self-contained dynamical system for $s_i$ and $h$. When making this substitution, we approximate the second term on the right-hand side of (5a), which describes the evolution of price over long-term horizons, by a positive constant representing the growth rate of the stock market.[15] We obtain the following equations:

$$\tau_i \dot{s}_i = -s_i + \tanh(\beta_1 s_i + \beta_2 h), \quad i = 1, 2, \ldots, N, \tag{6a}$$

$$\tau_h \dot{h} = -h + \tanh(\gamma \dot{s} + \delta + \kappa \xi_t), \tag{6b}$$

where $s = \frac{\sum \tau_i s_i}{\sum \tau_i}$ in accordance with (4), $\delta$ is a positive constant proportional to the stock market growth rate, $\gamma = \kappa_1 a_1$ and $\kappa$ is $\kappa_2$ renamed.

Equations (6) define a dynamical system of $N + 1$ mutually-coupled nonlinear equations. As we will see later, this coupling leads to interesting, nontrivial behaviors in the system, such as the emergence of self-sustained oscillations and their synchronization.

To develop further intuition about this system, we express it in the following approximate form (see Appendix A):

---

[15] Since $s \sim \sum \tau_i s_i$ and $|s_i| \leq 1$, the term $a_2(s - s_*)$ in (5a) is dominated by the sentiment of long-term investors, that is $s_i$ corresponding to large $\tau_i$. Also, equation (5b) implies that $s_i$ varies by $O(1)$ over $\tau_i$, i.e. the longer the investment horizon, the slower the sentiment variation. Therefore, $a_2(s - s_*)$ contributes to price development over the long term, e.g. months and years, which enables us to replace it in the leading order by a constant growth rate.

$$\tau_i \ddot{s}_i + G(s_i)\dot{s}_i + \frac{dU(s_i)}{ds_i} = F_i^c + F^e, \quad i = 1, 2, \ldots, N, \tag{7}$$

where $U(s_i)$, $G(s_i)$, $F_i^c$ and $F^e$ are given by equations (A3), (A4), (A5) and (A6), respectively.

Equations (7) govern the motion of $N$ particles (oscillators), each representing a network of investors characterized by horizon $\tau_i$, driven by the applied force. Interestingly, as follows from (7), $\tau_i$ takes on a meaning of the mass of the $i$-th particle in the sense that particles with small $\tau_i$ ("light" particles) are more sensitive to any force than particles with large $\tau_i$ ("heavy" particles). We can say that particles with small $\tau_i$ have smaller inertia than particles with large $\tau_i$.

The form of equations (7) allows their interpretation in terms of the particle's motion inside the potential well $U(s_i)$ in the presence of damping $-G\dot{s}_i$, influenced by the external force $F_i^c$ stemming from interaction between the particles (via common information flow $h$) and the external force $F^e$ due to the flow of exogenous news.

As follows from (A3), the shape of the potential well $U(s_i)$ is identical for all particles. Further, it depends only on two parameters, $\beta_1$ and $\delta$. In the case $N = 1$, Gusev et al. (2015) estimated $\beta_1 = 1.1$ and $\delta = 0.03$, which results in an asymmetric double-well shape of the potential (Figure 1).[16]

---

[16] As $\beta_1$ increases, the potential well undergoes a bifurcation from a single-well U-shape to a double-well W-shape at $\beta_1 = 1$. The potential is symmetric for $\delta = 0$; positive $\delta$ breaks the symmetry, making the part of the well where sentiment is positive deeper and the part where sentiment is negative more shallow.

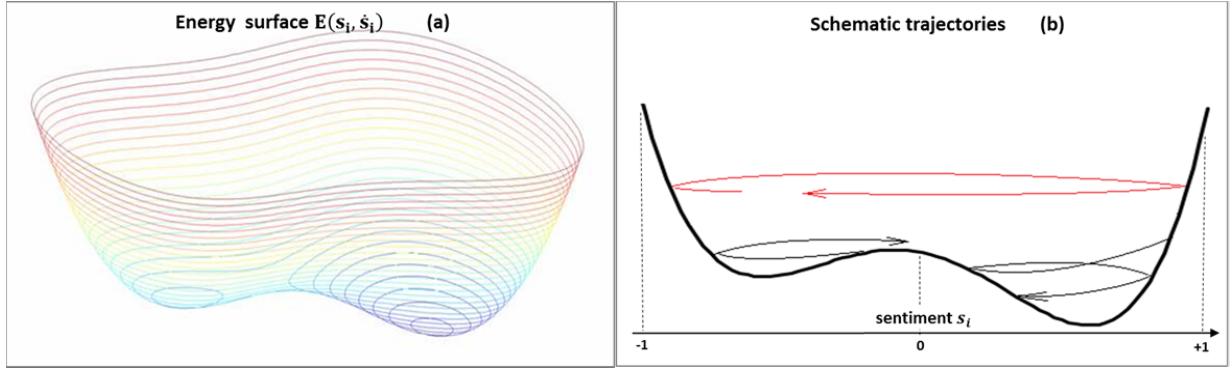

**Figure 1**: The profiles of the energy surface and the potential well corresponding to $\beta_1 \gtrsim 1$ and $0 < \delta \ll 1$. (a) The energy surface $E(s_i, \dot{s}_i)$ is shown as a function of $s_i$ and $\dot{s}_i$ in the space $(s_i, \dot{s}_i, E)$. The colors indicate energy levels, from low (blue) to high (red). (b) The potential well $U(s_i)$ is shown as a function of $s_i$. The equilibrium point at the cusp of the potential is the unstable saddle, while the equilibrium points at its minima can be stable or unstable nodes or stable or unstable foci, depending on the magnitude of feedback strength $\gamma$. The part of the well where sentiment is positive is deeper than the part where sentiment is negative for $\delta > 0$. Three typical trajectories are shown schematically in the well.

Figure (1a) depicts a surface corresponding to the kinetic and potential energy of the $i$-th particle as a function of $s_i$ and $\dot{s}_i$: $E(s_i, \dot{s}_i) = \frac{\tau_i}{2}\dot{s}_i^2 + U(s_i)$. Its cross-section by the plane $(E, s_i)$ gives the shape of the potential well and by the plane $(E, \dot{s}_i)$ gives the familiar parabolic profile of the kinetic energy. All trajectories lie on this energy surface. Figure (1b) depicts typical trajectories that we will discuss below.

If the impacts of damping $(-G\dot{s}_i)$, interaction $(F_i^c)$ and news $(F^e)$ were negligible, a particle would oscillate periodically in response to the restoring force $-dU/ds_i$ along the energy conserving trajectories, given by $E(s_i, \dot{s}_i) = $ constant, on horizontal planes.

Let us consider the impact of damping on a particle's motion. Damping, if not counteracted, causes the particle to lose energy, so that its path spirals down toward either the negative or positive stable equilibrium points located in the minima of $E(s_i, \dot{s}_i)$. Momentarily returning from this analogy to the real world, we can say that the interaction among investors within each peer group, subject to random idiosyncratic influences, compels the group's sentiment toward either a negative or positive equilibrium, where the consensus of opinion will be reached.

Price feedback adds a fascinating twist to this dynamic. It follows from (A4) that the damping coefficient $G$ is a function of the particle's position and is also dependent on several parameters, most notably the feedback strength $\gamma$. Interestingly, $G$ becomes negative in some regions on the $(s_i, \dot{s}_i)$-plane for $\gamma$ exceeding a certain critical value (equation A7), implying that for sufficiently strong feedback, damping begins to supply energy to the system instead of dissipating it. As a result, for large $\gamma$, some or all trajectories may converge to the limit cycle orbit where the supplied and dissipated energies compensate each other. This yields a potentially new state of dynamic equilibrium in which $s_i$ would exhibit self-sustaining, large-amplitude, periodic oscillations above the cusp of the energy surface between negative and positive sentiment values (the red trajectory in Figure 1b).

The critical value of $\gamma$ is roughly the same for all $s_i$ (equation A9). Therefore, for supercritical $\gamma$, the total sentiment would undergo the limit cycle oscillations, giving rise to the permanent regime of rallies and crashes, which contradicts the observed market behavior. Conversely, Gusev et al. (2015) showed for the case $N = 1$ that subcritical $\gamma$ leads to realistic market regimes. We should briefly inspect this case because in the absence of interaction ($F_i^c$), the heterogeneous model ($N > 1$) is qualitatively similar to the homogeneous model ($N = 1$) under the approximation (7).

The case $N = 1$ offers a simple portrait of trajectories as numerical solutions to equations (6) for $\xi_t = 0$ on the $(s, h)$-plane (Figure 2a). The distinct regimes illustrated schematically in Figure (1b) in the $(s, \dot{s})$-space are clearly visible here.[17] First, there are small-amplitude, decaying oscillations around the positive equilibrium point inside the deep well. Second, there are large-scale trajectories passing above the cusp of the potential, along which the particle can escape from one well into the other. Third, oscillations also occur inside the shallow well, where sentiment is negative, but as the equilibrium point there is unstable, the particle is quickly ejected onto the trajectories leading into the well in which sentiment is positive. Figure (2b), which depicts the empirical sentiment path traced by the US stock market during 1995-2015, confirms the existence of the above-described three types of sentiment motion.

Thus, we can presume that subcritical $\gamma$ permits realistic trajectories of sentiment evolution and so will apply subcritical values of $\gamma$ in the numerical and empirical analyses in the next sections.

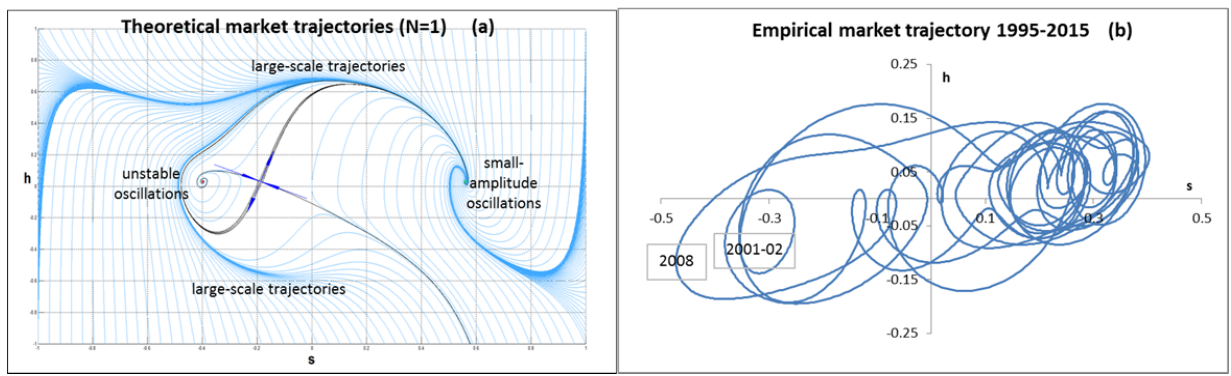

**Figure 2**: (a) The phase portrait on the $(s, h)$-plane for the homogeneous model ($N = 1$), showing an unstable focus in the negative-sentiment well (red asterisk), a stable focus in the positive-sentiment well (green asterisk) and large-scale trajectories crossing the wells (from Gusev et al.,

---

[17] The motion on the $(s, h)$-plane bears resemblance to the motion on the $(s, \dot{s})$-plane.

2015). (b) The phase portrait of the empirical market sentiment trajectory (1995-2015). To make this plot, the empirical time series of daily $h(t)$ and $s(h(t))$ have been obtained for $N = 1$ using the methodology outlined in Section 2.3 and then smoothed by a Fourier filter, removing harmonics with periods less than 100 business days. This path remained predominantly in the positive well, with only two excursions into the negative well during the bear markets of 2001-2002 and 2008.

Next, we examine the influence of the stochastic force $F^e$ generated by the flow of exogenous news (equation A6). This force acts to thrust a particle randomly from one trajectory to another, occasionally forcing it into a region which can lend the particle new dynamics, e.g. from the vicinity of the equilibrium points to the large-scale trajectories that traverse the well and vice versa. Thus, exogenous news flow plays a key role in market dynamics, being a random external force that may, from time to time, trigger changes in market regimes. Note that the asymmetry of the energy surface implies that a stronger force is needed to move a particle onto a path crossing from the deep (positive) well to the shallow (negative) well.

Additionally, owing to their lower inertia, "light" particles with small $\tau_i$ react more strongly to $F^e$ than "heavy" particles with large $\tau_i$. As a result, "light" particles can be expected to appear frequently on large-scale trajectories high on the energy surface, while "heavy" particles are likely to spend most of their time orbiting the equilibrium points at its bottom (Figure 1b).[18] This situation is relevant for the stock market since a greater volatility in sentiment is expected from short-term investors than from long-term investors.

---

[18] This analysis is relevant for the particles with $\tau_i \gtrsim \tau_h$. In Section 2.4 we will show that the "ultra-light" particles with $\tau_i \ll \tau_h$ possess no intrinsic dynamics, adjusting instead to the dynamics of "heavier" particles.

Finally, there are effects due to the force $F_i^c$ exerted on the $i$-th particle by the other particles (equation A5). Its action can be viewed through the prism of constraints imposed on the motion by the relations between each pair of particles in terms of their mutual positions and velocities that restrict the degrees of freedom of the motion.

We can write down equations for these constraints by observing that according to equation (6a) at any time all particles must share the same $h$ when moving along their paths on the energy surface. Accordingly, we invert (6a) to express $h$ as a function of $s_i$ and $\dot{s}_i$ and obtain an equation for the constraint $f_{ij}$ between the $i$-th and $j$-th particles:[19]

$$f_{ij} = f(s_i, \dot{s}_i; s_j, \dot{s}_j) = \frac{\operatorname{arctanh}(s_i + \tau_i \dot{s}_i) - \beta_1 s_i}{\operatorname{arctanh}(s_j + \tau_j \dot{s}_j) - \beta_1 s_j} = 1, \quad i = 1, 2, \ldots, N, \quad j = 1, 2, \ldots, N. \tag{8}$$

Equations (8) determine the relations between the sentiments and the rates of change in the sentiments of different investor groups due to mutual influences exerted by these groups on each other. These relations drive synchronization patterns, discussed in the next sections, plausibly resulting in self-similar behaviors, as well as other effects, in the market.[20]

Together, the above-described forces generate diverse and complex dynamics. For example, "light" particles may in response to negative news migrate from higher orbits in the well to orbits in the vicinity of the negative sentiment equilibrium at the well's bottom. According to (8), this change in the dynamic of "light" particles will require that "heavy" particles adapt their motion to synchronize frequencies and amplitudes. Should this dynamic persist, "heavy" particles, which make a major contribution to total sentiment (4), may cross from the positive well into the negative well, tipping overall sentiment in the market from positive to negative and, as a result, pressure market price downward. We will encounter this scenario of a bear market transition in numerical simulations and empirical analysis in the next sections.

---

[19] These constraint equations constitute $\frac{(N-1)N}{2}$ nontrivial first integrals of motion, out of which $N - 1$ are independent. The independent first integrals reduce the degrees of freedom of system (7) from $2N$ to $N + 1$, which matches the number of equations in the dynamical system (6).

[20] Synchronization is ubiquitous among the behaviors of interacting nonlinear oscillators: e.g. Pikovsky et al. (2001) provide an in-depth treatment of various synchronization effects in coupled oscillator systems.

At this point, we wish to remind the reader of the main purpose of this section, that is, to develop a conceptual understanding of the dynamics in model (6) prior to submitting it to the brute force of numerical simulations. We have therefore severely truncated this model to isolate the forces acting on a particle (i.e. an investor group) in equation (7) and explored the dynamic stemming from each force separately. The intuition developed here will aid in untangling the dynamics obtained in the next sections, with the caveat that these interpretations remain inexact because a specific dynamic of a particle, strictly speaking, can neither be completely attributed to one particular force, nor considered in isolation from other particles.

## 2.2. Numerical simulations

In Sections 2.2-2.3, we investigate system (6) numerically for $\tau_i \gtrsim \tau_h$, while the case where $\tau_i \ll \tau_h$ will be treated in Section 2.4. We use the estimate $\tau_h \sim 1$ day, which is consistent with the behavior of the autocorrelation of empirical $h(t)$, measured in Section 2.3, showing a fast decay of "memory" effects on the order of 1-3 business days.

We proceed, first, by considering only two groups of investors, with $\tau_1 = 1$ business day and $\tau_{15} = 15$ business days, to illustrate the effects discussed in the previous section. In terms of oscillator dynamics, the groups with $\tau_1$ and $\tau_{15}$ behave, respectively, as "light" and "heavy" particles on the energy surface.

Figure 3 depicts one simulation spanning 700 business days. The "light" particle undergoes a large-scale motion high in the potential well, covering the distance between extreme negative and extreme positive sentiment values in a 1-2 week timeframe. However, this particle can also get caught in a small-scale motion around the equilibrium points at the well's bottom, sometimes staying there for extended periods of time before it can escape (e.g. the intervals 120-170 days and 250-290 days in the positive well and the intervals 290-440 days and 610-690 days in the negative well). As discussed in the previous section, these transitions between large- and small-scale motions are triggered by the stochastic force exerted by exogenous news flow $\kappa \xi_t$.[21]

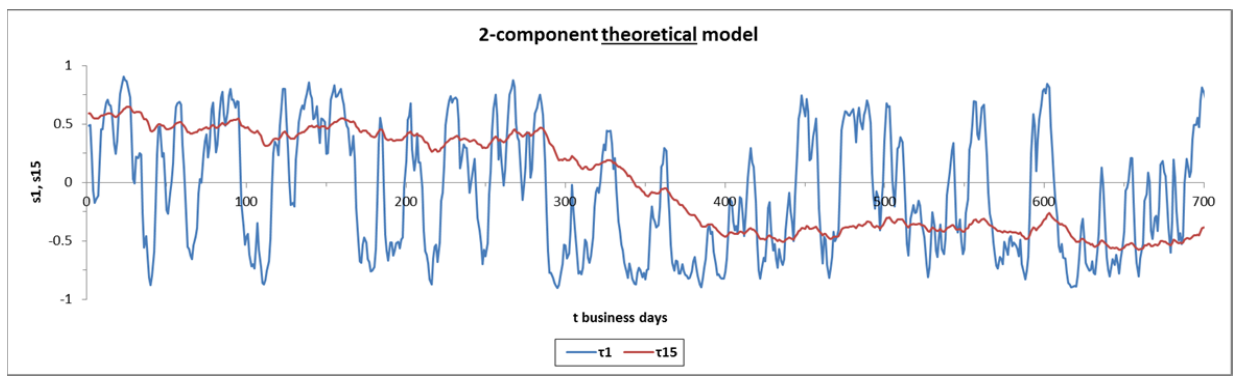

**Figure 3**: Sentiment evolution in the two-component theoretical model with $\tau_i = 1, 15$. Other parameters: $\beta_1 = 1.1, \beta_2 = 1.0, \delta = 0.02, \tau_h = 1, \gamma = 5$.

The "heavy" particle stays on the orbits near the bottom of the positive well during the first 300 days. Its motion is correlated with the motion of the "light" particle, such that its path shifts toward negative or positive values when the "light" particle is in the negative or positive sentiment region, respectively. Equation (8) attributes this behavior to the synchronization of the particles' dynamics. As a result, by observing the motion of one particle, we can deduce the motion of the other. For example, when the "light" particle remains sufficiently long as the solitary particle inside a well, the "heavy" particle will move from the well where it resides into the well in which the "light" particle is residing. Indeed, we observe that the "heavy" particle follows the "light" particle into the negative well in the interval 300-400 days. Visually it appears as if the "light" particle pulls the "heavy" particle across the wells. As we will see below, this is the basic characteristic of the cascade mechanism governing regime transitions between bull- and bear markets.

---

[21] On daily intervals, $\xi_t$ is modeled as normally-distributed white noise with zero mean and unit variance. However, we have chosen $\xi_t$ to have a small positive intraday autocorrelation on the assumption that news events are positively correlated on intraday time intervals; the autocorrelation is zero over the intervals of one day or longer.

Let us discuss the results of simulations in a more realistic model that consists of nine investor groups with $\tau_i = 1, 2, 3, 4, 11, 15, 19, 24, 28$ business days, which will be applied to design trading strategies in Section 3. Figure 4 shows sentiment evolution for four groups with $\tau_i = 1, 3, 11, 19$. The synchronicity among these groups is evident. For example, in the interval 750-800 days we can observe how the move of the group with $\tau_i = 1$ from the negative to positive well causes a similar move of the group with $\tau_i = 3$, followed by the group with $\tau_i = 4$ and then the rest of the groups, each with progressively smaller amplitude. Thus, it appears that regime transitions occur as the cascades propagating from "light" investors with small $\tau_i$ toward "heavy" investors with large $\tau_i$.[22]

---

[22] In this paper we do not consider the effect of a slowly varying $\beta_1$ on regime transitions, studied in detail by Gusev et al. (2015) for $N = 1$. These authors showed that $\beta_1$ slightly increased during the bull markets and slightly decreased during the bear markets (while remaining above unity), affecting the shape of the potential well and, therefore, the probability of regime transitions.

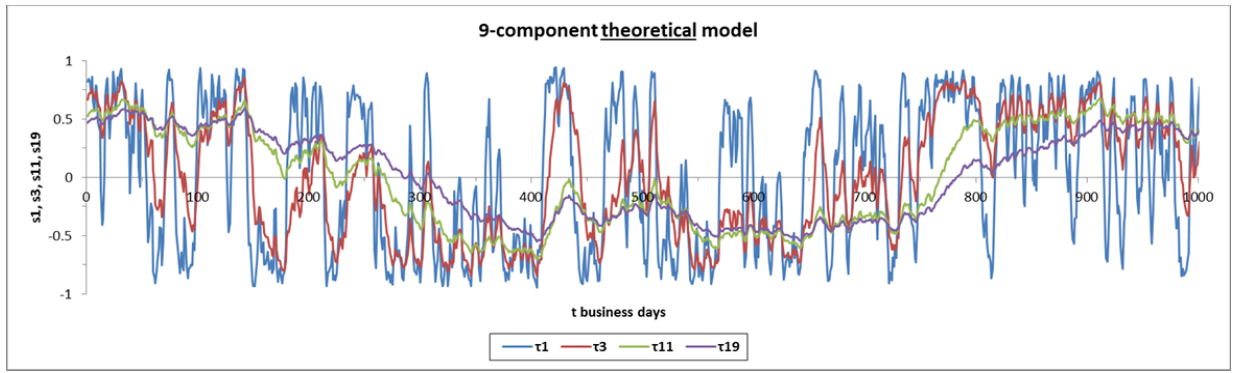

**Figure 4:** Sentiment evolution in the nine-component theoretical model with $\tau_i = 1, 2, 3, 4, 11, 15, 19, 24, 28$. Other parameters: $\beta_1 = 1.1, \beta_2 = 1.0, \delta = 0.02, \tau_h = 1, \gamma = 10$.

Another pattern discernable in this figure is that the particles form two groups with distinct dynamics: the first group with $\tau_i = 1, 3$ that follows a typical "light" particle dynamic and the second group with $\tau_i = 11, 19$ that behaves like a typical "heavy" particle. This separation implies a relatively sharp transition between the two dynamics as a function of the investment horizon $\tau_i$. Therefore, for a qualitative analysis, it seems justified to approximate interaction in the market as the interaction between two types of participants: volatile short-term investors who are sensitive to incoming information and relatively-static long-term investors whose views on the market are firmly established. Both groups are vital to market dynamics since the short-term investors are sufficiently "nimble" to initiate a change in market regime, while the long-term investors are sufficiently "massive" to actually effect the change.[23]

---

[23] Incidentally, the behavior of these two investor groups resembles that of the two types of investors ubiquitous in the market modeling literature: fundamental traders and systematic traders.

## 2.3. Empirical application

At this juncture we are ready to test the model with empirical data. For this purpose, we wish to measure $h$ for a relevant market and then calculate $s_i(h)$ and $p(h)$ from equations 5(b) and 5(a), respectively. We choose the US stock market as the object for this empirical study, given the news volume triggered by it, and select the S&P 500 Index as its proxy.

We remind the reader that $h$ has been defined in Section 1.1 as the ratio of the number of news items with positive market return expectations minus the number of news items with negative market return expectations over the total number of relevant news items. Thus, our objective is to capture in the general daily news flow the number of news releases that provide some indication of anticipated market movement: e.g. "… S&P 500 is likely to tumble amid worries over company profits…" or "… S&P 500 is expected to rise on upbeat jobs data…". We consider only the English language media and apply a rule-based parsing methodology outlined in Gusev et al. (2015) to daily news data for the period 1995-2015 retrieved from DJ/Factiva news archive. As a result, we obtain a time series of daily $h(t)$ over this period (Figure 5a).[24]

---

[24] A more detailed discussion about the measurement of $h$ can be found in Gusev et al. (2015). We only note here that these authors proposed to treat each news item as if it were a "sales pitch" aimed at investors to buy or sell the market. This allowed them to employ ideas from marketing research and put forward an argument that $h$ – information patterns that explicitly mention the direction of expected market movement – impacts investors most. In practice, news about current and recent market price changes, which make up the bulk of the study's relevant news volume, can also influence investors. As such, this information is included in the measurement of $h$, along with information concerning anticipated market movement.

To calculate sentiment, we use the nine-component model with $\tau_i = 1, 2, 3, 4, 11, 15, 19, 24, 28$ business days and $\beta_1 = 1.1$ and $\beta_2 = 1.0$, studied above. We substitute $h(t)$ into the corresponding equations (5b), solve them numerically for $s_i(t)$ and then obtain total empirical sentiment $s(t)$ from equation (4) (Figure 5b).

Next, we apply equation (5a) to calculate model price $p(t)$ from $s(t)$. To do that, we determine the coefficients in (5a) that minimize the mean-square deviation between $p(t)$ and the S&P 500 Index log prices. The resulting empirical model price $p(t)$ is shown in Figure 5c.

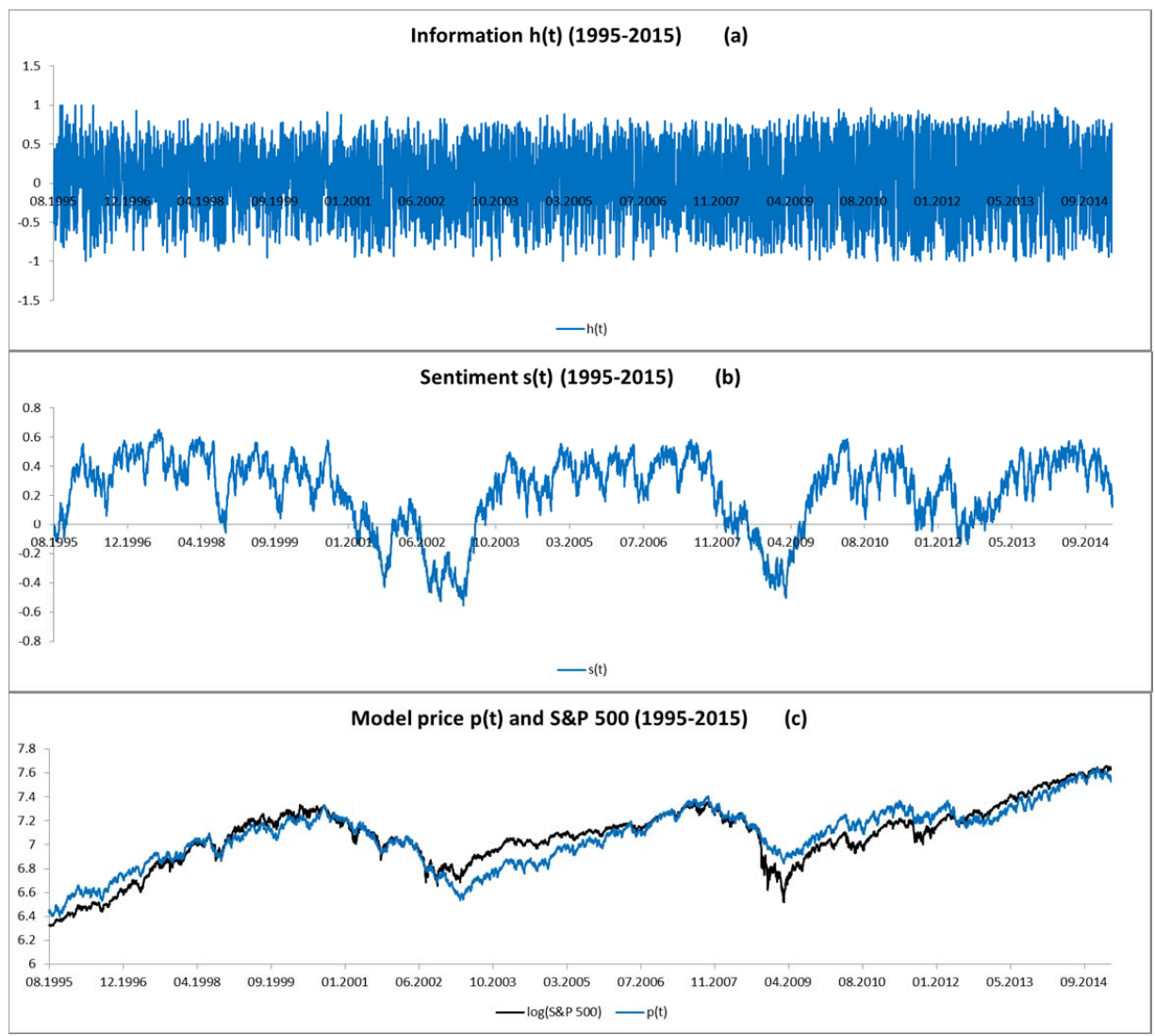

**Figure 5**: Daily time series of information $h(t)$, sentiment $s(t)$ and price $p(t)$ from 1995 to 2015. Sentiment $s(t)$ has been calculated from measured $h(t)$ using the nine-component model (5b) with fixed parameters $\tau_i = 1, 2, 3, 4, 11, 15, 19, 24, 28, \beta_1 = 1.1, \beta_2 = 1.0$. Price $p(t)$ has been calculated from $s(t)$ using the price formation equation (5a) with $a_1 = 0.356, a_2 = 0.003, s_* = 0.153$ and the integration constant equal to 6.455, estimated by least squares fitting.

The correlation between the daily model prices and the daily index log prices is over 92%. Thus, this heterogeneous model fits the market data better than the homogeneous model in Gusev et al. (2015), which is 83% correlated with actual prices. Such an accurate replication of the market price path is encouraging but it cannot resolve whether the model is predictive or not. We will test the predictability in Section 3, using trading strategies constructed upon this model.

Figure 6 shows the evolution of several $s_i(t)$ for the period 2005-2009, chosen to highlight the empirical behavior of sentiment during transition to- and from the bear market regime. These results are visually similar to the results of numerical simulation in Section 2.2. We particularly note the cascade mechanism of the market regime transitions and the distinct patterns in the behavior of the short-term and long-term investor groups, thus corroborating the main features of the model dynamics empirically.

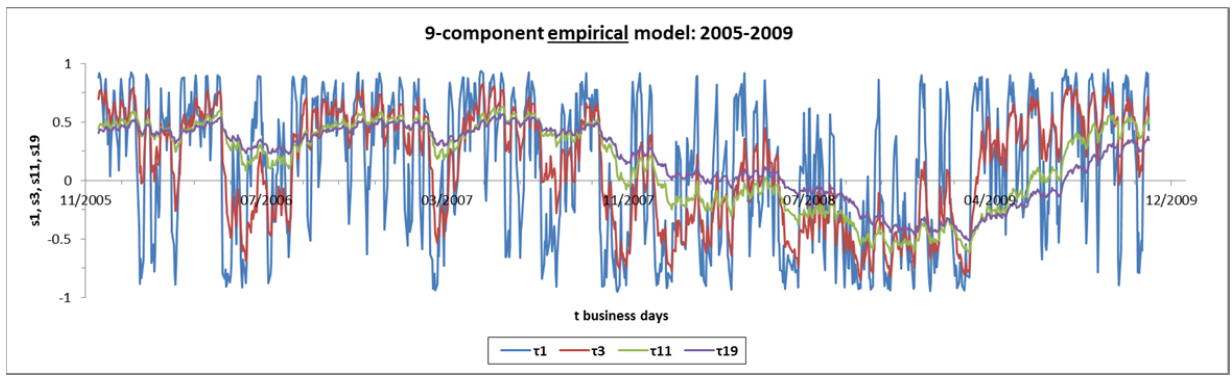

**Figure 6**: Sentiment evolution in the nine-component empirical model with $\tau_i = 1, 2, 3, 4, 11, 15, 19, 24, 28$. Other parameters: $\beta_1 = 1.1, \beta_2 = 1.0$.

## 2.4. Efficient market regime

In this section we show that in the leading order the dynamic of short-term investors decouples from the dynamics of investor groups with longer horizons, where short-term investors are defined as traders operating on timescales much shorter than $\tau_h \sim 1$ day. In particular, we will see that investment processes on these timescales are not involved in the feedback mechanism, but instead cause market price to adjust quickly to new information, contributing to market efficiency.

Let us consider a two-component [25] system (6) such that $\tau_1 \ll \tau_h \lesssim \tau_2$:

$$\tau_1 \dot{s}_1 = -s_1 + \tanh(\beta_1 s_1 + \beta_2 h), \tag{9a}$$

$$\tau_2 \dot{s}_2 = -s_2 + \tanh(\beta_1 s_2 + \beta_2 h), \tag{9b}$$

$$\tau_h \dot{h} = -h + \tanh(\bar{\gamma}(\tau_1 \dot{s}_1 + \tau_2 \dot{s}_2) + \delta + \kappa \xi_t), \tag{9c}$$

where $\bar{\gamma} = \frac{\gamma}{\tau_1 + \tau_2}$.

We first examine this system on timescales $\sim \tau_1$. It follows from (9a) that $s_1$ can change by $O(1)$ over $\tau_1$. Similarly, (9b,c) imply that $h$ and $s_2$ change respectively by $O\left(\frac{\tau_1}{\tau_h}\right) \ll 1$ and $O\left(\frac{\tau_1}{\tau_2}\right) \ll 1$ over $\tau_1$. Such a slow variation in $h$ and $s_2$ can in the first order be neglected, leading to

$$\tau_1 \dot{s}_1 = -s_1 + \tanh(\beta_1 s_1 + \beta_2 h), \tag{10a}$$

$$\dot{s}_2 = 0, \tag{10b}$$

$$\dot{h} = 0. \tag{10c}$$

---

[25] This result holds for model (6) with $N \geq 2$. For simplicity, we show its derivation in the case $N = 2$.

According to equations (10), both $h$ and $s_2$ remain approximately constant on timescales $\sim \tau_1$, over which $s_1$ converges from any initial position toward its equilibrium state ($\dot{s}_1 = 0$), given by

$$s_1 = \tanh(\beta s_1 + \beta_1 h). \tag{11a}$$

Next, we study system (9) on timescales $\sim \tau_h$ or longer. As viewed on these timescales, $s_1$ almost instantaneously ($\sim \tau_1 \ll \tau_h$) arrives at the position of equilibrium (11a). Consequently, sentiment $s_1$ behaves as if it were in a state of permanent equilibrium, while $h$ and $s_2$ evolve according to

$$\tau_2 \dot{s}_2 = -s_2 + \tanh(\beta_1 s_2 + \beta_2 h), \tag{11b}$$

$$\tau_h \dot{h} = -h + \tanh(\bar{\gamma} \tau_2 \dot{s}_2 + \delta + \kappa \xi_t). \tag{11c}$$

Therefore, system dynamics on these timescales are determined by equations (11b,c), whereas $s_1$ merely follows any changes in $h$ by moving along the equilibrium solution (11a), which is called the isocline. [26,27]

---

[26] Strictly speaking, although $s_1$ spends most of its time ($\sim \tau_h$) on the isocline, where its velocity is close to zero, it can also leave the isocline and move briefly ($\sim \tau_1$) along a trajectory in its vicinity (Figure 7). Therefore, $\dot{s}_1$ is nearly zero at all times, except for brief moments when the trajectory departs the isocline, so that the average contribution in (9c) due to $s_1$ is small as compared to $s_2$ and can be neglected.

[27] This approximation does not work in a system with $N = 1$, as there exist $\bar{\gamma}$ for which the term $\bar{\gamma} \tau_1 \dot{s}_1$ in (9c) cannot be neglected. As a result, for large $\bar{\gamma}$ the coupling between $s_1$ and $h$ can be strong enough to cause a limit cycle dynamic. The situation is different in systems with $N \geq 2$ that simulate market dynamics with a greater precision. There, $\tau_1 \dot{s}_1$ can in average be neglected in comparison with $\tau_i \dot{s}_i$ ($\tau_i \gtrsim \tau_h$) in (9c), so that the motion of $s_1$ is completely determined in this case by the dynamics between $h$ and $s_{i \neq 1}$.

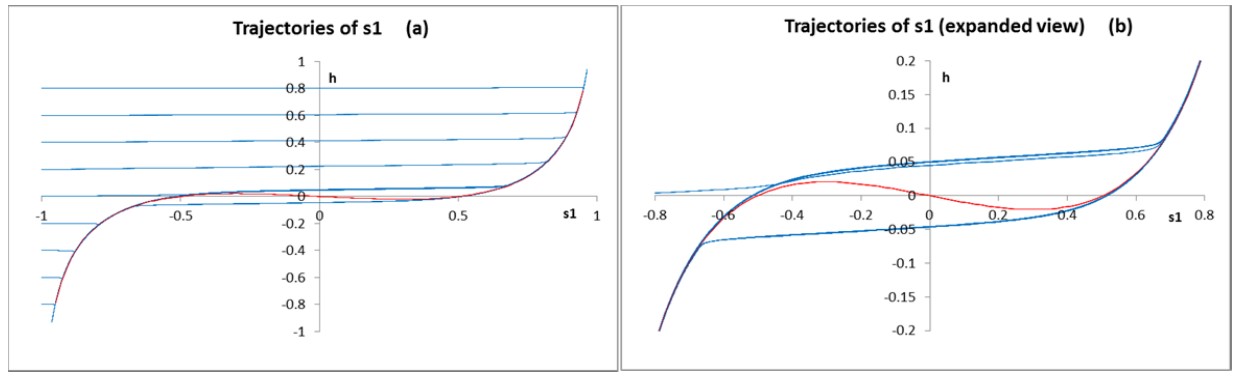

**Figure 7**: The isocline (red) and trajectories of $s_1$ (blue) for different initial conditions in system (9) with $\tau_1 = 0.01$ and $\tau_2 = 25$ business days. Sentiment $s_1$ falls on the isocline along the approximately horizontal lines: the motion occurring so fast that $h$ has little time to change. Sentiment $s_1$ continues to move along the isocline, following slowly evolving $h$. The segment of the isocline between its extrema is unstable, which causes sentiment to vacillate between the isocline's left and right branches. The overall motion consists of slow passages along the isocline and fast jumps between its branches, determined solely by the dynamics between $h$ and $s_2$.

Thus, we have shown that sentiment $s_1$, which develops on timescales $\tau_1 \ll 1$ day, decouples from the system's dynamics and does not participate in the sentiment-information feedback; instead, sentiment $s_1$ resides in a state of approximate equilibrium, adjusting instantaneously to changes in information flow $h$ and so driving corresponding changes in market price.[28] We therefore conclude that market efficiency persists on timescales much less than one day. Further, we can reasonably conjecture that intraday investment processes, generally, take place in a quasi-efficient market regime due to weak feedback, gradually giving way to the dynamics of mutually coupled information and sentiment over horizons longer than one day, which we study in this paper.

---

[28] This analytical result, which follows from equations (10) and (11), was also verified by direct numerical simulations. In addition, we note that equations (10) and (11) can be obtained by rescaling system (9) using a dimensionless time variable and then inspecting the leading-order balance on relevant timescales; we have chosen an informal derivation above for the sake of preserving the readability of this section.

## 3. News-based trading strategies

The previous section concluded that the stock market is efficient on short timescales $\tau_1 \ll \tau_h \sim 1$ day. This conclusion also sheds light on the mechanics of (intraday) news-based trading. Analysts take exogenous news flow $\xi_t$ as an input to generate and propagate information $h$ in time $\sim \tau_h$. Once $h$ is released, the short-term traders will move market price by $\Delta p \sim \Delta s_1$ in time $\sim \tau_1$, i.e. with a practically instantaneous effect. It follows that the objective of news-based trading is to capture this $\Delta p$ by estimating $h$ from $\xi_t$ before $h$ has been released.

Model (5) states that this same release of information $h$ will cause further changes in price, namely due to $\Delta s_2$, $\Delta s_3$, $\Delta s_4$ and so forth, that will be unfolding over days, weeks and months. In this section, we aim to verify this statement by designing and testing algorithmic trading strategies that can capture these longer-term impacts.

Our approach is as follows. In accordance with (5a) and (4), price changes are determined by the sentiments of investor groups with different investment horizons, which contribute to the formation of aggregate market sentiment on different timescales. Thus, if we extract the characteristic sentiment dynamic pertaining to each group, we can forecast price over various time horizons and implement trades based upon these forecasts.

We therefore wish to capture the characteristic dynamic of each investor group, while taking into account the influence of the other groups. Then, we apply this dynamic to extrapolate the future market position from its current position, given by empirically obtained $h$ and $s(h)$, and so generate return forecast over the relevant horizon. In terms of the motion in the $(N + 1)$-dimensional phase space $(h, s)$, this means investigating the characteristic behavior of the phase trajectory $(h(t), s(t))$ projected on the $(h, s_i)$-plane, subject to constraints imposed on $s_i$ by $s_{j \neq i}$.

In practice, the nine-component model (5) with $\tau_i = 1, 2, 3, 4, 11, 15, 19, 24, 28$ business days, studied above, has been applied to produce return forecasts over time horizons corresponding to the characteristic timescales in the model. These return forecasts can form the basis of a number of trading strategies, four of which, with different holding periods, are presented here. Specifically, we show one strategy based on the shortest forecast, two strategies based on different combinations of the equally-weighted forecasts and the last strategy based on the longest forecast. In the backtest results, the average holding periods of these strategies have been, respectively, around 9, 12, 25 and 45 business days (Table I).

Each strategy generates daily a buy-, sell- or hold signal on the SPDR S&P 500 ETF (Bloomberg ticker: SPY) ("SPY"), an exchange-traded fund tracking the S&P 500 Index, such that today's trading instruction is applied to the next day's opening price. The signal has no price input: it is based solely on the forecast derived from news. We emphasize that these strategies are merely crude prototypes, designed not for actual trading but to verify return predictability; as such, these strategies do not include position sizing and risk management.

We backtest these strategies over the period 1995-2015. Since the construction of strategies requires an in-sample period of roughly 2000 business days for parameter value selection, the backtested performance is reported for the out-sample period 2003-2015. We note that the strategies are not sensitive to the location of the in-sample period in the backtest interval and that in-sample and out-sample performance statistics are similar. The backtested results are compared with those of two benchmarks: a passive, long-only investment in SPY and an active, long-short strategy that combines a 5-day reversal and 250-day momentum strategies ("Mom-Rev") applied to SPY over this same period.

Figure (8) and Tables I and II show the cumulative returns, performance statistics and cross-correlations, respectively.

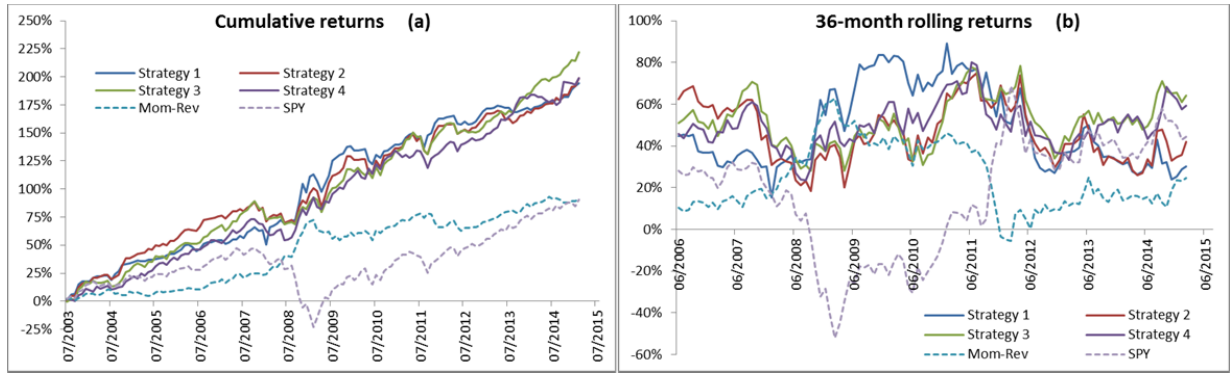

**Figure 8**: Performance graphs of four algorithmic news-based strategies applied to SPY, unadjusted for dividends, benchmarked against SPY and Mom-Rev (as defined above) during the out-sample period 2003-2015. The invested capital, as a base for daily P&L accruals in active strategies, was subject to a requirement, applied on the in-sample, that each strategy be on average fully invested during the periods for which trading signal was nonzero. No transaction costs were applied. Also, no interest and no funding costs were accrued on the under- and overinvested days, respectively. (a) Cumulative non-compounded monthly returns. (b) 3-year rolling returns.

**Table I: Statistics based on monthly returns**

|  | Strategy 1 | Strategy 2 | Strategy 3 | Strategy 4 | Mom-Rev | SPY |
|---|---|---|---|---|---|---|
| Mean return (%, p.a.) | 16.6 | 17.1 | 19.0 | 17.0 | 7.9 | 7.7 |
| Volatility (%, p.a.) | 15.2 | 15.5 | 13.9 | 13.9 | 9.4 | 14.6 |
| Max. drawdown (%, monthly) | -21.3 | -29.0 | -21.5 | -20.9 | -15.9 | -29.6 |
| 5%-VaR (%, monthly) | -5.4 | -7.1 | -5.9 | -5.3 | -3.9 | -8.6 |
| Gross exposure to SPY (%) | 61 | 72 | 96 | 100 | 42 | 100 |
| Alpha | 1.400 | 1.228 | 1.382 | 1.348 | 0.697 | 0.000 |
| Beta | 0.011 | 0.302 | 0.319 | 0.111 | -0.060 | 1.000 |
| Sharpe ratio* (p.a.) | 0.97 | 0.97 | 1.23 | 1.08 | 0.62 | 0.39 |
| Sortino ratio* (p.a.) | 0.91 | 0.76 | 1.15 | 1.23 | 0.56 | 0.32 |
| Holding period (bus. days) | 8.6 | 12.0 | 25.3 | 44.6 | 3.7 | n/a |

\* The risk free rate in the Sharpe ratio and the Sortino ratio is 2.0% p.a.

**Table II: Correlations (%) based on monthly returns**

|            | Strategy 1 | Strategy 2 | Strategy 3 | Strategy 4 | Mom-Rev | SPY |
|------------|------------|------------|------------|------------|---------|-----|
| Strategy 1 | 100        | 79         | 62         | 33         | 24      | 1   |
| Strategy 2 |            | 100        | 83         | 45         | 17      | 29  |
| Strategy 3 |            |            | 100        | 64         | 20      | 33  |
| Strategy 4 |            |            |            | 100        | 7       | 12  |
| Mom-Rev    |            |            |            |            | 100     | -9  |
| SPY        |            |            |            |            |         | 100 |

The pro-forma returns of all four news-based strategies have exceeded the returns of the equity index and the priced-based momentum-reversal strategy, on absolute- and risk-adjusted bases, and have also exhibited a relatively low correlation with these benchmarks on the 12-year out-sample period. Note that the lengths of the average holding periods of these strategies are substantially longer than that of the active benchmark. These results point toward return predictability on timescales longer than intraday and indicate that the model has, at least partially, captured this predictability.

It had been intended that this section would end with the sentence above. However, following the completion of this paper, we were able to substantially enhance the forecast precision in certain market regimes and, using these forecasts, develop an algorithmic news-based strategy sufficiently robust for implementation. We launched this trading strategy on October 23, 2015, with Interactive Brokers, an online broker, with the objective to test the model predictability in actual trading. We briefly describe this strategy, including the backtested results, in the following paragraphs.

The strategy has no price inputs, such as stop-losses, volatility-scaled exposure or other price-dependent features. The daily trading signal is derived solely from news and the principles set out in the present paper and is applied to trade SPY at the next day's opening price. The (pro-forma) average holding period is about 8 business days. The exposure to SPY is a function of the forecast probability, taking any values between -175% and +175%. In the backtest, the strategy displayed on average a 92% gross exposure to SPY.

Figure 9 shows the strategy's pro-forma cumulative returns and performance statistics, subject to an 0.80% p.a. cost drag, consistent with the transaction costs currently incurred by this strategy in actual trading. The strategy has been tracked against the passive benchmark (SPY) and the active benchmark (Mom-Rev), defined above. The out-sample results are as follows: The strategy is more aggressive than the benchmarks, since it has exhibited a 25.1% annual volatility vs SPY's 14.6% and Mom-Rev's 9.5%. However, its downside risk is below or comparable with that of the benchmarks, as measured by the monthly 5%-VaR of -4.7% vs SPY's -8.2% and Mom-Rev's -4.0% and the worst monthly drawdown of -14.4% vs SPY's -29.6% and Mom-Rev's -15.9%. The strategy has performed on a risk-adjusted basis better than both benchmarks, with the Sharpe ratio of 1.07 vs SPY's 0.31 and Mom-Rev's 0.47 and, especially, with the Sortino ratio of 2.52 vs SPY's 0.26 and Mom-Rev's 0.43. It has demonstrated an equally strong outperformance on an absolute return basis, as is evident from the graphs. These results lend further credence to the market model studied here and warrant further research on its applications to market prediction, including actual trading.

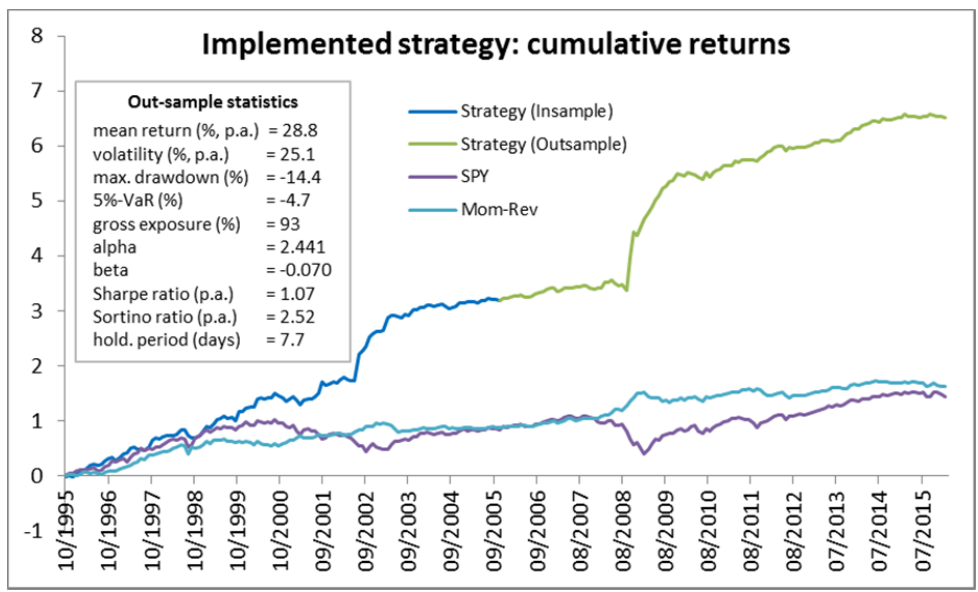

**Figure 9**: Backtested performance of the implemented news-based strategy (traded since 10/2015) benchmarked against SPY and Mom-Rev, defined above, during 10/1995-01/2016. The strategy signal and the active benchmark (Mom-Rev) signal are applied to the next day's opening price of SPY, unadjusted for dividends. The figure offers the graphs of cumulative non-compounded monthly returns, as well as the summary out-sample statistics for the news-based strategy in the embedded table (the Sharpe and Sortino ratios are calculated with a 2.0% p.a. risk-free rate).

## 4. Discussion

The starting point for this paper was the agent-based stock market model in Gusev et al. (2015). The model consists of analysts[29], who extract relevant information from price changes and exogenous news, and investors, who apply this information to trade. The interaction mode among the agents is assumed to be all-to-all to derive the model equations in analytic form as a dynamical system governing the evolution of the mutually-coupled endogenous "macroscopic" variables: market price ($p$), investor sentiment ($s$) and information ($h$) supplied by analysts (equations 1).

This homogeneous interaction topology, assumed in the model, is instrumental for identifying the basic mechanisms that drive market dynamics; however it also makes the model insufficiently fine-grained for testing return predictability. An introduction of a more complex topology is not a straightforward task for the lack of obvious choices and because of the sensitivity of model's properties to these choices..

---

[29] This term is applied in a collective sense, comprising financial analysts, newspaper journalists, market commentators, finance bloggers and other participants who communicate their market views through mass media.

We sidestepped this problem by arranging investors in peer networks, according to investment horizons, on the assumption that interaction among peers is strongest, obtaining in the limiting case the all-to-all interaction within each peer network and zero outside interaction (Section 1.2). Despite the absence of interaction across the networks, each still can impact the others by contributing to the common information flow that affects all networks in equal measure. This phenomenological approach has enabled us to derive a heterogeneous model with $N$ investor networks or groups (equations 5), which is sufficiently realistic to be applied for return prediction and yet simple enough to be expressed in analytic form.

In particular, this model demonstrates that with respect to processing information the market behaves quasi-efficiently on intraday timescales (Section 2.4) and inefficiently on timescales longer than one day (Sections 2.2-2.3). The model equations reveal that it is price feedback that enforces market inefficiency by coupling the endogenous variables. We have shown that feedback is negligible on the intraday scale, but is important over longer time horizons, where it contributes to leading-order dynamics.

The situation where a system exhibits different behaviors on different scales is not unusual in nature. Fluid dynamics provides an instructive example. In fluids, inertia plays a central role on large scales, while viscous damping is dominant on small scales.[30] As a result, the large-scale dynamics and the small-scale dynamics are fundamentally different: inertia induces a nonlinear endogenous dynamic at macro scales, whereas at micro scales inertia is so small that velocities in a fluid's flow adjust immediately to exogenous changes, leading to a state of adiabatic equilibrium.

---

[30] For example, we swim by using water's resistance to create momentum; however this strategy would fail if we were the size of bacteria: for microorganisms, water appears as viscous as honey for humans, forcing them to evolve unique propulsion techniques, such as corkscrew-like locomotion mechanisms among others.

The above example presents a useful analogy for market dynamics as the market also evolves on many (time)scales, driven by participants with various investment horizons. Indeed, equation (7) implies that the investment horizon $\tau_i$ is analogous to the mass of the $i$-th investor group's sentiment in the context of sentiment dynamics. Accordingly, the contribution of inertia to dynamics on short timescales is negligible because the mass of the sentiment of relevant investor groups is very small: these investors react so fast as to move prices almost instantaneously in response to new information, leading to an (adiabatic) equilibrium regime on these timescales (Section 2.4). On the contrary, inertia cannot be neglected on longer timescales, which results in effective interaction between investors and analysts in the model, yielding complex dynamics characterized by nonlinear feedback (Sections 2.2-2.3).

The intraday market efficiency does not imply the lack of short-term trading opportunities. In fact, whereas price adjusts instantaneously, information $h$ is released by analysts on average on the scale $\tau_h \sim 1$ day. This delay, which can likely be attributed to information processing (e.g. gathering, aggregation, analysis, editing) and distribution frequency, creates a window of opportunity for intraday trading between the occurrence of a news event (e.g. the release of an earnings report) and its reflection in $h$.

This short-term price reaction to information released by analysts is incomplete because the overall market sentiment also includes the sentiments of investor groups with longer investment horizons, which can influence the mid- and long-term price evolution. Equations (6) state that a change in information will cause changes in sentiment on many different timescales and that these changes will in turn cause changes in information – creating a feedback loop. This complex multi-scale interplay between information and sentiment is the generator of the variety in market behavior, including self-similar variation patterns briefly explained as a synchronization effect in Sections 2.1-2.3.

In addition, the market is subject to the impact of exogenous news flow that acts as an external stochastic driving force (equations 5). However, as discussed above, on long timescales the market acquires inertia and with it a resistance to change in direction. As a result, market behavior can be predictable in situations where inertia outweighs noise. A test of this predictability has been carried out in Section 3.

Our approach to return prediction is based on principles similar to those of weather forecasting, i.e. combining theoretical models and empirical measurements. We have obtained the empirical time series of information (Section 2.3) and applied model (5) to forecast sentiment and price and, based on these forecasts, develop the prototypes of trading strategies (Section 3). The backtested results, compared to passive and active benchmarks, suggest that market forecasting on the above-described principles functions with a precision sufficient for the development and implementation of successful news-based trading strategies operating over horizons ranging from days to months.

We have sought to produce a market model that is sufficiently sophisticated to both replicate past performance and predict future returns, while being tractable to highlight the essential mechanisms underlying market dynamics. We fully realize that the range of processes occurring in the market is substantially broader than those captured by this model. For example, the model does not include fundamental traders (who apply financial analysis) and systematic traders (who use price data). We note, however, that the analysts in the model perform analogous functions, so that, in the first order, the impacts due to these two types of investors have been taken into account. In any case, this is work in progress – modeling necessarily proceeds from simple to complex as the grasp of underlying mechanisms improves – the objective of which is to advance the current knowledge of market dynamics and provide a basis for further modeling efforts.

We would like to conclude this section by referencing Farmer (2001) who argued that agent-based models had the potential to give practitioners better tools to predict markets but noted that "the advent of practical agent-based models is still at least several years away". To the best of our knowledge, the news-driven model of stock market dynamics developed in the present paper is the first such practical agent-based model.

## 5. Conclusion

In this paper we have theoretically and empirically investigated stock market return predictability on various time horizons. In particular, we introduced a news-driven model with heterogeneous investors and, using this model, developed and backtested purely news-based, algorithmic trading strategies. In the course of this study we have reached the following conclusions:

1. There exist two characteristic timescales of stock market dynamics. Over time horizons shorter than one day, the market behaves quasi-efficiently with respect to processing information. On time horizons longer than one day, the market becomes inefficient.
2. This informational inefficiency is caused by a feedback loop, which acts on timescales longer than one day, interconnecting information, opinion and price and giving rise to fundamentally nonlinear overall dynamics.
3. On these timescales, the relevant model for market dynamics is a dynamical system governing the evolution of mutually-coupled information, opinion and price, driven by exogenous news.
4. According to this model, the sentiments of investor groups with different investment horizons collectively form aggregate investor opinion that determines a price dynamic, which in turn influences information flow acting on all groups participating in the market.
5. This common information flow provides a link through which the sentiments of investor groups are mutually coupled. As such, information induces self-similar dynamics among investor groups on multiple timescales through synchronization, leading to complex self-similar patterns observable in market behavior.

6. These investor groups form two classes characterized by distinct dynamics. The first class contains investors with horizons less than one week. Their average sentiments are volatile, typically changing from negative to positive or vice versa in the timeframe from one to four weeks. The second class consists of investors with horizons exceeding one week. Their average sentiments normally undergo small-amplitude oscillations around either a positive or negative equilibrium, where the consensus of opinion is reached.

7. The regime change between bull- and bear markets takes place when investors with long investment horizons transit from one sentiment equilibrium to the other. This transition occurs as a cascade, whereby investors with longer horizons follow, one-by-one, investors with shorter horizons.

8. The backtested results of the prototypes of trading strategies, designed by blending a theoretical agent-based model (dynamical system) with empirical observations (news data) for trading over time horizons that range from days to months, suggest that the stock market dynamics are to a certain extent predictable.

9. Having provided theoretical and empirical evidence for market predictability in this paper, it is our future research objective to test this predictability in actual trading. In the end of Section 3, we briefly reported the pro-forma risk-return characteristics of the news-based strategy that we have recently launched to initiate such a test. This strategy is currently trading only the US stock market. Our next steps include extending it to other equity markets and, potentially, also to commodity markets.

**Acknowledgments**

We are grateful to LGT Capital Partners for partially funding this project and to Dow Jones & Company for providing access to the Factiva.com news archive. We would like to thank a number of people who provided their help over the course of this work: John Orthwein for editing this paper and contributing ideas on its readability; Ed King and Maxim Zhilyaev for helping with data retrieval

**Appendix A: Approximation of the dynamical system**

Here we express the dynamical system (6) as a system of forced, coupled, nonlinear oscillators.[31] We differentiate equation (6a) with respect to time and use equation (6b) to obtain

$$\tau_i \ddot{s}_i = \Phi(s_i, \dot{s}_i, \dot{s}, \xi_t)$$

$$= -\dot{s}_i + (1 - (s_i + \tau_i \dot{s}_i)^2)\left(\beta_1 \dot{s}_i + \frac{\beta_1}{\tau_h} s_i - \frac{1}{\tau_h} \operatorname{arctanh}(s_i + \tau_i \dot{s}_i)\right)$$

$$+ (1 - (s_i + \tau_i \dot{s}_i)^2)\frac{\beta_2}{\tau_h} \tanh(\gamma \dot{s} + \delta + \kappa \xi_t), \quad i = 1, 2, \dots, N, \quad (A1)$$

where $s = \frac{\sum \tau_i s_i}{\sum \tau_i}$ in accordance with (4).

These equations govern the motion of $N$ oscillators – that is $N$ particles with the coordinates $s_i$ and the velocities $\dot{s}_i$, subjected to the force $\Phi(s_i, \dot{s}_i, \dot{s}, \xi_t)$. Note that $\tau_i$ is analogous to the mass of the $i$-th particle in the sense that the impact of a force on the particles with small $\tau_i$ ("light" particles) is greater than on the particles with large $\tau_i$ ("heavy" particles). In other words, "light" particles have small inertia and "heavy" particles have large inertia.

---

[31] We follow the steps of a similar derivation for $N = 1$ in Gusev et al. (2015) (Appendix C). That appendix also provides a detailed analysis of the phase portrait geometry, including the bifurcations of equilibrium points and the formation of a stable limit cycle.

The first two terms in $\Phi(s_i, \dot{s}_i, \dot{s}, \xi_t)$ contain the restoring and damping force components responsible for autonomous dynamics. The third term describes the force originating from the $i$-th sentiment component feedback ($\sim \gamma \tau_i \dot{s}_i$ and $\sim \delta$) and the external forces exerted by the other particles ($\sim \gamma \sum_{j \neq i} \tau_j \dot{s}_j$) and by the flow of exogenous news ($\sim \kappa \xi_t$) in the argument of the hyperbolic tangent. Being dependent on position and velocity, these forces vary along a particle's trajectory.

For illustration purposes, we expand $\Phi(s_i, \dot{s}_i, \dot{s}, \xi_t)$ into a truncated Taylor series to separate the above-mentioned force components and write equation (A1) in a canonical form:

$$\tau_i \ddot{s}_i + G(s_i) \dot{s}_i + \frac{dU(s_i)}{ds_i} = F_i^c + F^e, \quad i = 1, 2, \ldots, N. \tag{A2}$$

In this equation, $U(s_i)$ has the meaning of a potential and is given with the precision up to a constant by

$$U(s_i) = \frac{1}{\tau_h} \left( \frac{\beta_1 - \frac{2}{3}}{4} s_i^4 - \frac{\beta_1 - 1}{2} s_i^2 - \beta_2 \delta s_i \right); \tag{A3}$$

$G(s_i)$ has the meaning of a damping coefficient and is given by

$$G(s_i) = \left(1 - \beta_1 - \beta_2 \bar{\gamma} \frac{\tau_i}{\tau_h} + \frac{\tau_i}{\tau_h}\right) + 2\beta_2 \frac{\tau_i}{\tau_h} \delta s_i + \left(\beta_1 + \beta_2 \bar{\gamma} \frac{\tau_i}{\tau_h} + 2(\beta_1 - 1) \frac{\tau_i}{\tau_h}\right) s_i^2; \tag{A4}$$

where

$$\bar{\gamma} = \frac{\gamma}{\sum \tau_i};$$

$F_i^c$ has the meaning of an external force exerted by the other particles and is given by

$$F_i^c = \frac{\beta_2 \bar{\gamma}}{\tau_h} \sum_{j \neq i} \tau_j \dot{s}_j; \tag{A5}$$

and $F^e$ has the meaning of an external force due to the flow of exogenous news and is given by

$$F^e = \frac{\beta_2}{\tau_h}\kappa\xi_t. \quad (A6)$$

As such, equation (A2) describes the motion of a particle inside an asymmetric double-well potential well (A3) for $\beta_1 > 1$ (see Figure 1) in the presence of nonlinear damping (A4), driven by the forces generated through interaction between particles (A5) and through the impact of exogenous news (A6). Note that the feedback force in (A1), proportional to $\bar{\gamma}\tau_i\dot{s}_i$ and $\delta$, has been incorporated into both the potential force (only the component $\sim\delta$) and the damping force on the left-hand side of (A2).

To obtain equations (A2)-(A6), we have truncated the Taylor series of $\Phi(s_i, \dot{s}_i, \dot{s}, \xi_t)$ at terms above cubic in $s_i$, linear in $\dot{s}_i$ and linear in $\delta$ and have kept only the leading terms in the expressions for the forces $F_i^c$ and $F^e$. Consequently, these equations are, strictly speaking, only valid in the region where $|s_i| \ll 1$ and $|\dot{s}_i| \ll 1$. However, we expect that the formula for the potential $U(s_i)$, which does not contain the heavily truncated terms $\sim \dot{s}_i$, holds reasonably well for all sentiment values ($|s_i| \leq 1$) within the relevant range of parameter values, that is $\beta_1 \sim 1, \beta_2 \sim 1$ and $\delta \ll 1$.

As follows from (A4), the damping coefficient $G(s_i)$ is negative if

$$\bar{\gamma} > \bar{\gamma}_c(s_i, \tau_i) = \frac{\left(1 - \beta_1 + \frac{\tau_i}{\tau_h}\right) + 2\beta_2\frac{\tau_i}{\tau_h}\delta s_i + \left(\beta_1 + 2(\beta_1 - 1)\frac{\tau_i}{\tau_h}\right)s_i^2}{\beta_2\frac{\tau_i}{\tau_h}(1 - s_i^2)}. \quad (A7)$$

Condition (A7) means that for sufficiently large $\bar{\gamma}$ there are regions where energy in the system is amplified (negative damping), pointing toward the possibility of a limit cycle. Because this condition has been derived for $|s_i| \ll 1$ and $\delta \ll 1$, we can in the leading order neglect the terms $\sim \delta s_i$ and $\sim s_i^2$ to obtain

$$\gamma > \gamma_c(\tau_i) = \frac{1 - (\beta_1 - 1)\frac{\tau_h}{\tau_i}}{\beta_2} \sum \tau_i. \qquad (A8)$$

Since $\beta_1 \sim 1$ (we use $\beta_1 = 1.1$), the second term in the numerator in (A8) is much smaller than unity for particles with $\tau_i \geq \tau_h$ (we set $\tau_h = 1$ day) and can be neglected. This means that $\gamma_c$ has approximately the same value for all investors with investment horizons equal to or longer than one day, given by

$$\gamma_c = \frac{1}{\beta_2} \sum \tau_i. \qquad (A9)$$